# Application of the Monte-Carlo refractive index matching (MCRIM) technique to the determination of the absolute light yield of a calcium molybdate scintillator

V. Alenkov[1], O. A. Buzanov[1], N. Khanbekov[2], V. N. Kornoukhov[2*], H. Kraus[3], V. B. Mikhailik[3, 4**], V. A. Shuvaeva[5],

[1] JSC FOMOS-Materials, 107023, Moscow, Russia

[2] Institute of Theoretical and Experimental Physics, 117218, Moscow, Russia

[3] Department of Physics, University of Oxford, Keble Road, Oxford OX1 3RH, United Kingdom

[4] Diamond Light Source Ltd, Harwell Science and Innovation Campus, Didcot, OX11 0DE, United Kingdom

[5] Institute of Physics, Southern Federal University, 194, Stachki Str, Rostov-on-Don, 344090, Russia

## Abstract

The use of $^{40}Ca^{100}MoO$ in experimental searches for neutrinoless double beta decay (0νDBD) relies on knowledge of fundamental scintillation properties of the material. In this work we determine the absolute light yield of calcium molybdate using Monte-Carlo refractive index matching technique (MCRIM). The MCRIM technique is a combination of experiment and simulations that allows the absolute light yield of scintillators to be determined by taking into account effects of refraction, scattering and absorption in the material. The light collection efficiency of the scintillator-detector assembly was simulated using the ZEMAX ray-tracing software. By tuning the optical parameters of the scintillation crystal, a model was derived that gives good agreement with the experimental results. It is shown that the light collection efficiency of scintillators increases with transmittance and scattering due to an enhanced probability for photons to escape the crystal volume. Using MCRIM, the absolute light yield for the $^{40}Ca^{100}MoO_4$ scintillator was found to be 7.5±1.2 ph/keV at room temperature. Comparative measurements using a $CaWO_4$ scintillator as a reference show good agreement with this result. In that way, the study demonstrated the potential of the MCRIM technique as a tool for quantitative characterization of scintillation materials.

*-corresponding authors: e-mail: kornoukhov@itep.ru (V.N. Kornoukhov), tel +7 (499) 789-66-29

**-corresponding author e-mail: v.mikhailik@physics.ox.ac.uk (V.B.Mikhailik), tel. +44(0)1865-273459

## 1. Introduction

Understanding the fundamental properties of neutrinos further, provides strong motivation for experimental searches for neutrinoless double beta decay (0νDBD) of some even-even nuclei. The extremely low rate of 0νDBD events poses a significant challenge for distinguishing the signal from background due to natural radioactivity. The most efficient method for addressing this problem is the active discrimination of events using two different parameters of the detector response. Three techniques are of practical interest in this regard. Discrimination of scintillation events using pulse shape analysis has proven to be a robust technique permitting to achieve fairly competitive limits in 0νDBD search experiments [1]. The possibility of event discrimination using the pulse shape of signals from cryogenic bolometer has been demonstrated recently [2]. Finally, a very favourable method is the simultaneous detection of scintillation and phonons in cryogenic detectors, offering excellent sensitivity [3-6]. Due to natural abundance (9.63%) and its Q-value (3034 keV) for 0νDBD $^{100}$Mo is considered to be a very attractive nucleus for such experiments. Significant abundance of $^{100}$Mo permits enrichment at reasonable cost, while the high Q-value is particularly helpful for minimizing the background in the region of interest. These factors in particular prompted recently an interest in Mo-based scintillators for use in cryogenic rare event search experiments.

High scintillation efficiency is a primary selection criterion for a scintillation material to be used in cryogenic rare event search experiments and among the family of Mo-based inorganic oxides, $CaMoO_4$ is the one exhibiting a fairly high scintillation response at low temperatures [7-9]. Therefore, $CaMoO_4$ has been proposed as a target material for a large-scale experimental search for 0νDBD [10]. This prompted extensive research aiming at the optimization of scintillation properties of calcium molybdate [11-13]. Recently, we succeeded in the development of a technology for the production of large high-quality scintillation crystals from isotopically enriched materials with low intrinsic radioactivity [14]. The scintillation properties of the $^{40}Ca^{100}MoO_4$ crystals were evaluated by comparing them with a reference $CaMoO_4$ scintillator [9] but no absolute light yield was reported. Therefore, the aim of this work was to determine the absolute light yield of a $^{40}Ca^{100}MoO_4$ single crystal.

## 2. Background of the problem

The intrinsic or absolute light yield $Y$ is defined as the number of photons $N_0$ produced in a scintillation crystal per deposited energy $E$ of a gamma quantum:

$$Y = \frac{N_0}{E} \tag{1}$$

The measured number of photons $N$ is related to $N_0$ through the light collection efficiency $\eta$ (see below for details) and the emission-weighted detector efficiency $\varepsilon_\lambda$ as:

$$N = N_0 \eta \varepsilon_\lambda. \tag{2}$$

The emission-weighted detector efficiency takes into account the variation of the detector quantum efficiency over the emission spectrum of the scintillator:

$$\varepsilon_\lambda = \frac{\int \varepsilon(\lambda)s(\lambda)d\lambda}{\int s(\lambda)d\lambda}, \qquad (3)$$

where $\varepsilon = \varepsilon(\lambda)$ and $s(\lambda)$ are the wavelength-dependent quantum efficiency and emission intensity.

The main problem with such measurements is how to achieve an accurate determination of $\eta$. The light collection efficiency represents a complex convolution of the crystal's optical properties (absorption, scattering, refraction, surface conditions) and the experiment's geometry (crystal shape, size, coupling to detector, etc.). To reduce the uncertainty associated with $\eta$ the optimum would be to make this parameter close to unity. There are several practical approaches that allow increasing light collection substantially by reducing light trapping and self-absorption. The first task is achieved by using optical coupling (grease) and by covering the side surfaces with a diffusive reflector (Teflon tape) while the second is realized by using for measurements thin (~1 mm) samples [15-18]. It is common practice to assume that for such an experimental setting the light trapping and self-absorption are negligible; the final results are corrected only for the reflectivity of Teflon tape (0.95-0.98). Convergence of the results obtained for quite a few scintillators within an error of ±10% (see [18]) shows that the assumption regarding light collection are fairly reasonable. However, it has also been noticed that for inorganic scintillators with a large refractive index there is strong variation of the light yield with sample thickness [17, 19] and a larger than expected discrepancy between different measurements [18]. This is very likely due to unaccounted light losses as the light trapping tends to be more pronounced in materials with a large refractive index.

Therefore, for characterisation of such scintillators, it is necessary to quantify the light collection efficiency $\eta$. This task can be achieved by using Monte-Carlo simulations if the input parameters are known with sufficient precision, but usually this is not a case. Furthermore, some of the parameters are difficult to determine or they may vary during the course of the measurement. For example, the biggest challenge is the modelling of diffusive surfaces as this requires knowledge of the complex scattering distribution function of the surface. Lack of this information significantly reduces the accuracy of the simulations and hence the estimate of the light yield. On the other hand, ray-tracing is quite reliable if surface effects are eliminated. Recently, the Monte Carlo refractive index matching (MCRIM) method was devised by Wahl et al [20]. By excluding uncontrollable parameters and minimising the impact of all difficulties, the authors developed a combination of Monte-Carlo modelling with experiment, allowing derivation of the scattering characteristics of a crystal. This in turn permits calculation of the light collection efficiency $\eta$. The MCRIM is therefore a very promising method to aid the measurements of absolute light yield and the purpose of this study was to apply this technique to the $^{40}Ca^{100}MoO_4$ scintillator.

### 3. Method

The key consideration in MCRIM is to ensure that the experiment can be reproduced by Monte-Carlo simulation. To achieve good accuracy of such a simulation it is imperative to minimise the number of uncontrollable parameters influencing light collection $\eta$. This is achieved by removing from the system any dependencies due to possible variations of surfaces (all surfaces are deemed to be polished, no reflectors are used). For this case the function describing the light collection efficiency $\eta$ contains only two unknown parameters:

the absorption and scattering coefficients $\alpha_{abs}$ and $\alpha_{scat}$ respectively. Other parameters, i.e. emission wavelength, refraction indices and dimensions, are well defined. Measurements of transmittance $T$, which is related to $\alpha_{abs}$ through Beer's law

$$T = \exp(-\alpha_{abs} d) \quad , \tag{4}$$

where $d$ is the path length through the crystal, reduces the number of unknown parameters to one. Subsequently, Eq. (2) now contains two unknown parameters $\eta$ and $N_0$. As mentioned above, the light collection efficiency is a complex function of many parameters but only one ($\alpha_{scat}$) remains unknown. The second parameter $N_0$ is a constant number that can be excluded from the analysis by taking the ratio of two different measurements of $N$.

$$R_{1/2} = \frac{N_0 \eta_1 \varepsilon_\lambda}{N_0 \eta_2 \varepsilon_\lambda} = \frac{\eta_1}{\eta_1} \quad . \tag{5}$$

It is essential for these measurements to differ in a way that is reproducible in Monte Carlo simulations and that they provide information on $\alpha_{scat}$. Another element of MCRIM is to change only one variable of the experimental setup, specifically, to fill the gap between the crystal and scintillator with optical media of different refractive indices $n_{gap}$. The change in light collection efficiency due to changing $n_{gap}$ causes a corresponding change in the number of photons which can escape trapping without affecting the scattering coefficient $\alpha_{scat}$. By iterating the value of $\alpha_{scat}$ the Monte Carlo simulation aims to match the value of $R_{1/2}$ with that obtained in the experiment. When $\alpha_{scat}$ is established it is possible to work backwards and determine the collection efficiency $\eta$ for any of the performed experiments and, finally, to calculate $N_0$ using equation (2).

### 4. Experiment

In this study we used a crystal made of a sample of $^{40}Ca^{100}MoO_4$ scintillator with dimensions $10.0 \times 10.0 \times 10.0$ mm$^3$ that was previously described in [14]. All surfaces of the crystal were polished to a roughness of 50 nm. The transmittance of the crystal measured at the maximum of the luminescence band of calcium molybdate (540 nm) is 76%. When corrected for multiple reflection [21] this gives a transmittance $T = 96\%$ and an absorption coefficient $\alpha_{abs} = 0.039$ cm$^{-1}$. The refractive index of calcium molybdate used in this calculation is $n = 2.01$ [22].

The setup used for the measurements of the light yield is essentially identical to that described in [20]. It consists of a 59.5 keV γ-ray source ($^{241}$Am), a $^{40}Ca^{100}MoO_4$ crystal scintillator and a PMT, (model 5124A, ET Enterprises), all placed in a light tight box. The crystal was positioned on the window of the PMT and the photons created through scintillation were detected in photon counting mode using the multi-photon counting technique [23]. A significant advantage of this technique is the ability to distinguish individual photons resulting from scintillation events in slow scintillators [24]. This gives directly the number of photons per scintillation event, and eliminates the need for a potentially error-prone calibration procedure that depends on the accurate determination of the single electron response [16, 18]. The spectral dependence of the quantum efficiency of the PMT was taken from data supplied by the manufacturer. The relative error of this parameter is established as ±5% due to statistical variation between different devices and errors in the measurements of this characteristic [16-18]. The emission spectrum of $CaMoO_4$ is taken from our previous spectroscopic studies of this material [8]. The total error in the calculation of the emission-weighted detector efficiency $\varepsilon_\lambda$ is assessed to be ±10 % [16]; it results from adding in quadrature the quantum efficiency error and errors in spectroscopic measurements.

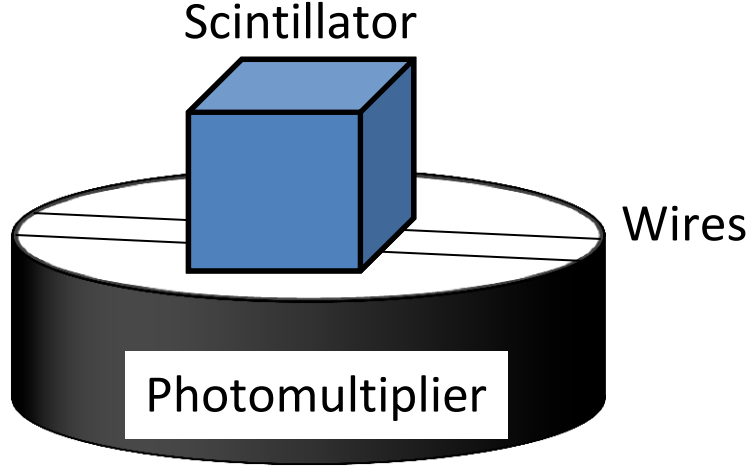

Fig. 1: Experimental setup showing the arrangement of the PMT, scintillator and a pair of wires of Ø0.05 mm, used for controlling the gap thickness.

A gap of controlled thickness (0.05 mm) was created by placing two supporting wires between the crystal and the PMT window (Fig. 1). Such a gap eliminates uncertainty due to uneven contact and gives a possibility for introducing in a controlled fashion a medium with refractive index $n_{gap}$ other than that of air. The measurements of the light yield of calcium molybdate were performed for two coupling geometries: i) with the empty gap ($n_{gap} = n_{air} = 1.00$) and ii) gap filled with Cargille optical gel 0607 with $n_{gap} = n_{gel} = 1.46$. Four data sets were collected for 10 minutes each in these configurations. To verify the effect of systematic error the setup was dismantled and assembled back for each measurement. The experimental error was obtained from the repeated independent measurements within one configuration of experimental setup.

### 5. MCRIM simulations

To simulate light transport in the calcium molybdate scintillator ─ detector setup, the ray-tracing software ZEMAX was used [25]. A model was created to reproduce the experimental setup shown in Fig. 1. All simulations were carried out in non-sequential mode, meaning that there is no predefined sequence of surfaces which rays that are being traced must hit. The trajectories of rays are determined solely by the physical positions and properties of the objects as well as the directions of the rays. Thus, the rays may hit any part of any object, and also may hit the same object multiple times. This allows total internal reflection to be accounted for. An advantage of ZEMAX over other packages is that it takes into consideration polarisation effects and ray splitting at the interfaces.

The source was simulated as a volume object with dimensions 0.001 mm less than the geometrical sizes of scintillation crystal to eliminate uncertainty at the surface. The optical parameters for simulation where taken from the literature or measurements as explained in sections 3 and 4. The entrance window of the PMT was modelled as a 28 mm × 28 mm × 1 mm plate with circular aperture 28 mm made of borosilicate glass BK7 with refractive index 1.52. A gap of 0.05 mm was introduced between the crystal and the PMT window. Switching between two coupling geometries was represented as a change of refractive index in the volume of the gap ($n_{gap} = n_{air} = 1.00$ or $n_{gap} = n_{gel} = 1.46$). For the calculation of light collection efficiency the detector was modelled as ideal absorber attached to the surface of the window.

In each case the 100,000 rays randomly distributed across the volume of the crystal were traced. The result of the simulations is the fraction of the total energy generated by the source object (scintillator) that reaches the detector. This number represents the light collection efficiency $\eta$ of the setup. The ray tracing was carried out for different scattering coefficients $\alpha_{scat}$. The variation of this parameter produced the set of simulation data for two coupling geometries (gap filled with air or optical gel) to be used later for comparison with the experimental results.

To verify the validity of the model and results produced by ZEMAX, simulations were performed for simple configurations that had analytical solutions for light collection. It has been shown that the amount of energy emitted by one plane of a rectangular parallelepiped with perfectly polished surfaces is equal to [26]:

$$\eta = \frac{1}{2}(1 - \frac{\sqrt{n^2-1}}{n}), \qquad (6)$$

where $n = n_{crystal}/n_{medium}$ is the relative index of refraction of the crystal with respect to the surrounding medium. Substituting values of $n$ for air and gel as surrounding medium (2.01 and 1.38) gives the light collection efficiency for the two geometries as equal to 6.6 and 15.6%, respectively. The simulations in ZEMAX carried out under the assumption of 100% transmission (no absorption or scattering) resulted in identical numbers within the error limit of less than 1%.

### 6. Determination of the absolute light yield

#### 6.1. Measurements of light output for $^{40}Ca^{100}MoO_4$

Figure 2 shows the typical histograms of number of photons per scintillation event due to 59.5 keV γ-rays measured for two coupling geometries. A value for the light output is obtained as position of peak centroid after a Gaussian fit of a band in a histogram of photon distribution per scintillation event. The measured light yield and error calculated from the repetition of four independent experiments is $N_{air}$=14.3±0.2 and $N_{gel}$=24.5±0.4 photons for the configuration with empty gap and gap filled with optical gel, respectively. The error represents the uncertainty in the peak position and not the width of the photon distribution.

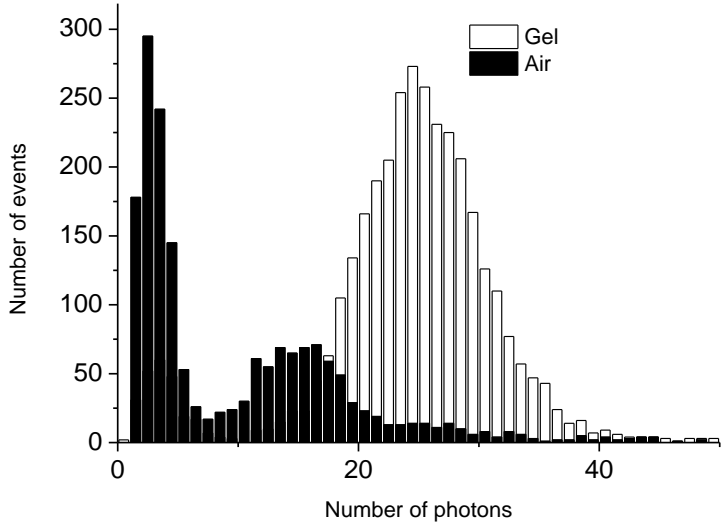

Fig. 2: Histograms of distribution of detected photons from scintillation events excited in $^{40}Ca^{100}MoO_4$ by 59.5 keV γ-rays from $^{241}Am$. The plots are given for two coupling geometries between crystal and detector: air gap and gel.

*6.2. Simulation of light collection efficiency*

The ratio of the light yields measured in two experiments is $R_{air/gel}$ = 0.583±0.012. As mentioned above, this parameter is independent of the absolute light yield of the crystal and gives a ratio of light collection efficiencies for two coupling geometries. These geometries were modelled by ZEMAX: the light collection efficiency $\eta$ was calculated for different scattering coefficients $\alpha_{scat}$ varying between 0.01 and 2 cm$^{-1}$ (see Fig. 2).

To assess the effect of transmittance on $\eta$, simulations were carried out for the nominal transmittance of $T$ =96.0% and also for $T$ =93.0% and 98.6%. The latter two values correspond to the limiting cases in the variation of transmittance of the $^{40}Ca^{100}MoO_4$ crystal over the 400-650 nm wavelength range when measured in different crystal orientations. Inspection of the curves displayed in Fig. 3 instantly confirms earlier empirical observations that the light collection efficiency increases with transmittance and bulk scattering; both parameters enhance the probability of a photon escaping the crystal. However, the effect of scattering on the light collection efficiency is more complex. For the coupling geometry with the air gap, when $\alpha_{scat}$ exceeds 0.5 cm$^{-1}$, $\eta$ behaves asymptotically, exhibiting little change. In the case of scintillator to detector coupling with gel, the light collection efficiency shows a maximum at around 0.3-0.4 cm$^{-1}$ and then gradually decreases.

Having calculated dependences of $\eta = f(\alpha_{scat})$ for both experimental geometries, the ratio $R_{air/gel} = \eta_{air}/\eta_{gel}$ is obtained straightforwardly (see Fig. 4). By comparing the results of the simulation and the experimental data the value for the scattering coefficient can be derived as $\alpha_{scat}$ = 0.54±0.10 cm$^{-1}$. Through Fig. 3 this parameter allows inferring values of the light collection efficiency as 25% and 43% for the detection geometry with air and gel coupling, respectively.

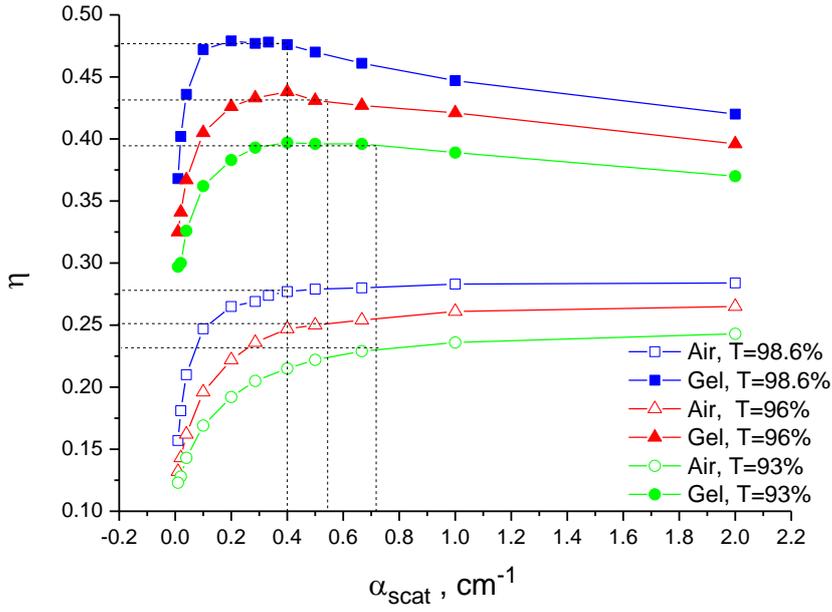

Fig.3. Light collection efficiency as a function of scattering coefficient calculated for two coupling geometries (gap filled with air and gel) and three values of crystal transmittance.

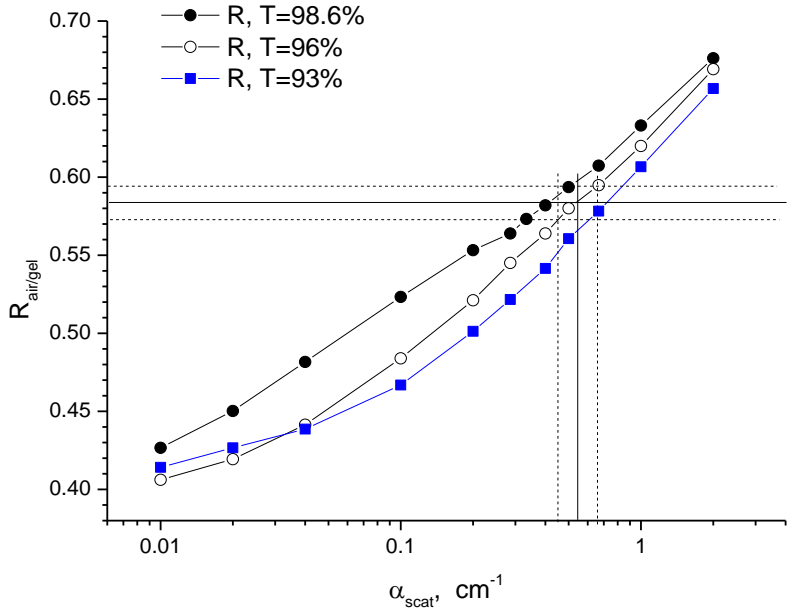

Fig. 4: Ratio of light collection efficiencies $R_{air/gel} = \eta_{air}/\eta_{gel}$ vs. scattering coefficient. The experimental value with associated error ($R_{air/gel}$=0.583±0.012) is marked by horizontal lines. Their intersection with the curve $R_{air/gel} = f(\alpha_{scat})$, obtained for $T$ =96.0 %, gives the value of scattering coefficient for the measured $^{40}Ca^{100}MoO_4$ crystal ($\alpha_{scat}$=0.54±0.10 cm$^{-1}$).

Given the importance of numerical information on light collection efficiency for further analysis the first question to ask is: how realistic are the results derived from these simulations? To answer this we shall refer to the relevant results of earlier investigations. In the classical study of the absolute light yield of scintillators Holl et al estimated that for a polished sample of a BGO scintillator coupled to a detector with gel without reflective wrapping, $\eta$ is about 60% [15]. It is worth mentioning that the samples used in [15] were of small thickness with small height-to-base aspect ratio that is beneficial for high light collection [16]. To make the comparison more relevant, the light collection efficiency of a thin 10×10×1 mm$^3$ sample of CaMoO$_4$ was simulated, and that gave $\eta$=63%. Next, in the original paper on the MCRIM technique [20] the light collection efficiency of a polished CaWO$_4$ crystal was calculated for a coupling geometry with an air gap. A value of $\eta$=15.2 % can be deduced from the quoted number of 3.8 %, corrected for the quantum efficiency of the detector (25%). Recently, a value of light collection efficiency $\eta$=41% was reported for a polished, un-coated LSO-Ce scintillator coupled to the detector with gel [27]. Altogether these results are in reasonable agreement especially when taking into account that they are obtained for samples with very different optical characteristics and through using different and independent simulation techniques.

### 6.3. Determination of the absolute light yield of $^{40}Ca^{100}MoO_4$

Once the light collection efficiency is deduced, the absolute light yield of the $^{40}Ca^{100}MoO_4$ scintillator under examination can be calculated for both geometries from equations (1) and (2):

$$Y_1 = \frac{N_{air}}{\varepsilon_\lambda \eta_{air} E} = \frac{14.3}{0.127 \times 0.25 \times 59.5} = 7.5 \, \text{ph/keV}$$

$$Y_2 = \frac{N_{gel}}{\varepsilon_\lambda \eta_{gel} E} = \frac{24.5}{0.127 \times 0.43 \times 59.5} = 7.5 \, \text{ph/keV}$$

It has been highlighted already that $\eta$ experiences noticeable changes with transmittance (see Fig. 3). To assess errors due to the uncertainty in this parameter the light collection efficiency was derived for the limiting values of transmission, 98.6 and 93.0%, that where inferred as mentioned above. This provides the range of $\eta$ values that are used to calculate the possible variation of the absolute light yield of the $^{40}Ca^{100}MoO_4$ scintillator (see Table 1).

Table 1. Variation of light collection efficiency of a $^{40}Ca^{100}MoO_4$ scintillator with transmittance $T$.

| $T$, % | $\alpha_{abs}$, cm$^{-1}$ | $\alpha_{scat}$, cm$^{-1}$ | $\eta_{air}$, % | $\eta_{gel}$, % |
|---|---|---|---|---|
| 98.6 | 0.014 | 0.42 | 28 | 48 |
| 96.0 | 0.039 | 0.54 | 25 | 43 |
| 93.0 | 0.072 | 0.73 | 23 | 40 |

Analysis of the results compiled in Table 1 allows making a conservative estimate of the relative error in the assessment of $\eta$ as ±12%. By taking into account the errors of the calculation of the emission-weighted detector efficiency (±10%) and the error of the measurements of number of photons (±1.6%) the total error when evaluating the absolute light yield is ±16%. Thus, the absolute light yield of $^{40}Ca^{100}MoO_4$ at room temperature is concluded to be 7.5±1.2 ph/keV.

*6.4. Relative method*

The absolute light yield of scintillation crystals is very often measured through relative methods in which a comparison with a scintillator of known light yield is made. In such a case the value of the unknown light yield is determined by comparing the number of photons produced at a given energy and correcting for the emission-weighted quantum efficiency of the detector and light collection efficiency. Comparative measurements of the light yield of calcium molybdate were carried out relative to $CaWO_4$ with an absolute light yield of 16.2±1.6 keV$^{-1}$ [28]. This scintillation crystal has been earlier characterised very comprehensively using the MCRIM method [20]. Using the known values for the absorption ($\alpha_{scat}$ = 0.065 cm$^{-1}$) and scattering coefficients ($\alpha_{scat}$ = 0.061 cm$^{-1}$) a light collection efficiency of $\eta$ =16.2 % was calculated for the coupling geometry with the air gap. This is very close to the value of $\eta$ =15.2 % quoted in [20] as mentioned above. The emission-weighted detector efficiency is 0.221. The measured light yield of the $CaWO_4$ scintillator sample in this coupling geometry was found to be 40.1±0.4 photons. The light yield of the $CaMoO_4$ scintillator is calculated using the relation:

$$\frac{Y_{CaMoO4}}{Y_{CaWO4}} = \frac{N_{CaMoO4}\varepsilon_{CaWO4}\eta_{CaWO4}}{N_{CaWO4}\varepsilon_{CaMoO4}\eta_{CaMoO4}} \quad (7)$$

Substituting the numerical values into this equation gives

$$Y_{CaMoO4} = \frac{14.3 \times 0.221 \times 0.162}{40.1 \times 0.127 \times 0.23} \times 16.2 = 7.1 \pm 1.7 \text{ ph/keV}$$

Evidently the values of absolute light yield when using two methods agree very well within the error limit. The exercise also highlights the fact that a simple comparison of two scintillators under the assumption of identical collection efficiency would give a rather underestimated value of 5.6 ph/keV. Such comparison is very common practice and it may be the cause of significant discrepancies between some results.

*6.5. Discussion*

The light yield of calcium molybdate has been studied earlier in a few works (see Table 2) and it is essential to discuss them in view of present findings. It can be seen that the results of relative measurements of the light yield fluctuate very significantly [7, 11, 12]. However, they fall outside this discussion as there is no practical way for converting these data into absolute values because of a large uncertainty in the light collection efficiency of the relevant measurements.

Table 2. Measured values of the light yield of $CaMoO_4$ crystals at room temperature.

| Relative LY, % | Absolute LY, ph/keV | Method | Reference |
| --- | --- | --- | --- |
| 20% (CsI-Tl) | | Relative to CsI-Tl | 7 |
| 4% (CsI-Tl) | | Relative to CsI-Tl | 11 |
| 36% (CaWO$_4$) | | Relative to CaWO$_4$ | 12 |
| 9% (NaI-Tl) | | Relative to NaI-Tl | |

|   | 4.9±0.59 | Avalanche photodiode + corrections | 13 |
|---|---|---|---|
|   | 3.0 | Photomultiplier + corrections | 29 |
|   | 8.9±3.5 | Relative to $CaWO_4$ | 9 |
|   | 7.5±1.2 | Photomultiplier + corrections | This work |

Up to date the absolute light yield of $CaMoO_4$ has been reported in three papers. The experimental results obtained by Kim et al [13], using avalanche photodiode are fairly reliable as this detector can be calibrated very accurately. However, the empirical correction factor applied by the authors to account for the light collection efficiency (90%) may be optimistic. As mentioned in the introduction this might be justified for materials with a not very large refractive index (<1.9) and thin samples where self-absorption due to light trapping is fairly low. For highly refractive materials the light collection efficiency should take these effects into account. The simulations were performed for a $CaMoO_4$ scintillator in geometry [14] using optical parameters obtained in the present study. The value of $\eta$ for the cubic scintillator in reflective wrapping with gel coupling to a detector is found to be 56%. After correction of the value of 3.5 ph/keV measured in [13] for $\eta$ and the detector efficiency (80%) the absolute light yield is found to be 7.8 ph/keV, which is very close to the result of the present study. The light yield reported by Vasiliev et al [29] was obtained using a formula that relates energy resolution to the statistical uncertainty of the detected signal. However, the contribution due to the intrinsic resolution of the crystal (can be as large as that due to the statistical fluctuation of the photomultiplier response [28]) was not included, leading to incompatible results. Mikhailik et al [9] used the relative method with a $CaWO_4$ reference crystal and obtained an adequate value of the light yield for $CaMoO_4$ (8.9±3.5 ph/keV) under the assumption of identical light collection efficiency. In this experiment a thin 1 mm sample is placed inside a cryostat and the scintillation light is collected over a small solid angle. The light collection efficiency is controlled mainly by the setup geometry. The optical properties of the crystals have a smaller effect as opposed to the conventional geometry of light yield measurements where the detector is in contact with the scintillator. Thus, it can be concluded that the value of the absolute light yield obtained in this study is compatible with the results of earlier works.

### 7. Conclusions

The determination of absolute light yield of scintillators relies on exactitude in evaluation of the light collection efficiency $\eta$ of a particular experimental setup. The MCRIM technique was developed by Wahl et al [20] as a combination of experiments and simulations that allows to account for the effect of optical properties of a material (refraction index, transmittance and scattering) on light collection. In this work we demonstrated how the MCRIM technique can be deployed to facilitate the determination of the absolute light yield of $^{40}Ca^{100}MoO_4$ scintillators. The refraction index and transmittance are obtained from independent optical experiments while the scattering coefficient is inferred by fitting to the results of the light yield measurements of the scintillator in two coupling geometries of the detector. Introducing different transparent media, namely, air or gel, in the gap between the scintillation crystal and detector changes the measured light yield, and that can be simulated by ray-tracing using a unique set of optical parameters for the crystal. Obtained in such way, the optical parameters are then used to calculate the light collection efficiency of the setup and consequently the absolute light yield of the scintillator under examination.

We used the ZEMAX ray-tracing software to carry out simulation of the light collection efficiency of the scintillator-detector assembly. By modelling the experimental setup it is shown that the light collection efficiency increases with transmittance and scattering within the crystal due to an enhanced probability for photons to escape. Both parameters have a significant effect and therefore must be accounted for in the light yield determinations.

Using MCRIM, a value for the scattering coefficient for the $^{40}$Ca$^{100}$MoO$_4$ scintillator with given transmittance $T$ = 96% is estimated to be $\alpha_{scat}$ = 0.54±0.10 cm$^{-1}$. The absolute light yield was calculated for two coupling geometries of the experiment (without and with gel filling of gap) resulting in the same value of 7.5±1.2 ph/keV at room temperature. To verify this finding we carried out comparative measurements using a CaWO$_4$ scintillator as a reference and demonstrated that results agree very well.

The simulations highlighted the significance of light trapping and self-absorption in crystals with high refractive index that leads to a noticeable reduction of the light collection efficiency. This is very often ignored, resulting in an over-estimated correction factor for the light collection. The examples relevant for this study are discussed and it is shown that by making the appropriate correction it is possible to achieve good consistency between the results of present and earlier works on absolute light yield of calcium molybdate. Overall this proves that the simulations underpinning the method give accurate predictions and MCRIM technique is a useful tool for the determination of fundamental characteristics of scintillation materials.

### Acknowledgment

This study was supported in part by Federal Science and Innovations Agency of Russian Federation (Federal Aiming Program State contract 16.523.11.30130 and by the Royal Society, London, UK.

### References

1. F. A. Danevich, A. S. Georgadze, V. V. Kobychev et al., Search for 2 beta decay of cadmium and tungstate isotopes: Final results of the Solotvyna experiment, Phys. Rev. C 68 (2003) 035501.
2. C. Arnaboldi, C. Brofferio, O. Cremonesi et al., A novel technique of particle identification with bolometric detectors, Astropart. Phys. 34 (2011) 797-804.
3. C. Arnaboldi, J.W. Beeman, O. Cremonesi et al., CdWO$_4$ scintillating bolometer for Double Beta Decay: Light and Heat anticorrelation, light yield and quenching factors, Astropart. Phys., 34 (2010) 143-150.
4. L. Gironi, C. Arnaboldi, J. W. Beeman et al., Performance of ZnMoO$_4$ crystal as cryogenic scintillating bolometer to search for double beta decay of molybdenum, J. Instrum. 5 (2011) P11007.
5. C. Arnaboldi, S. Capelli, O. Cremonesi, L. Gironi, M. Pavan, G. Pessina, S. Pirro. Characterization of ZnSe scintillating bolometers for Double Beta Decay, Astropart. Phys. 34 (2011) 344-353.
6. S .J. Lee, J. H. Choi, F. A. Danevich, et al. The development of a cryogenic detector with CaMoO$_4$ crystals for neutrinoless double beta decay search, Astropart. Phys. 34 (2011) 732-737.

7. S. B. Mikhrin, A. N. Mishin, A. S. Potapov, et al, X-ray excited luminescence of some molybdates. Nucl. Instr.Meth. A 486 (2002) 297.
8. V. B. Mikhailik and H. Kraus, Performance of scintillation materials at cryogenic temperatures, Phys. Stat. Sol. B 247 (2010) 1583.
9. V. B. Mikhailik, S. Henry, H. Kraus and I. Solskii Temperature dependence of $CaMoO_4$ scintillation properties, Nucl. Instr. Meth. A 583 (2007) 350.
10. H. J. Kim et al., A search for the 0-neutrino double beta decay with the $CaMoO_4$ scintillator, in: Proceedings of New Views in Particle Physics (VIETNAM'2004), August 5–11, 2004, p. 449.
11. S. Belogurov, V. Komoukhov, A. Annenkov et al., $CaMoO_4$ scintillation crystal for the search of Mo-100 double beta decay, IEEE Trans. Nucl. Sci. 52 (2005) 1131-1135.
12. A. N. Annenkov, O. A. Buzanov, F. A. Danevich, et al., Development of $CaMoO_4$ crystal scintillators for a double beta decay experiment with $^{100}$Mo, Nucl. Instr. and Meth. A 584 (2008) 334-345.
13. H. J. Kim, A. N. Annenkov, R. S. Boiko, et al., Neutrino-less double beta decay experiment using $Ca^{100}MoO_4$ scintillation crystal. IEEE Trans. Nucl. Sci. 57 (2010) 1475-1480.
14. V. V. Alenkov, O. A. Buzanov, N. Khanbekov et al., Growth and characterization of isotopically enriched $^{40}Ca^{100}MoO_4$ single crystals for rare event search experiments, Cryst. Res. Thechn. 46, (2011) 1223-1228.
15. I. Holl, E. Lorenz, and G. Mergas, A measurement of the light yield of common inorganic scintillators, IEEE Trans. Nucl. Sci., 35, (1988) 105–109.
16. M. Moszynski, M. Kapusta, M. Mayhugh, et al., Absolute light output of scintillators, NIEEE Trans. Nucl Sci. 44 (1997) 1052 – 1061.
17. M. Gierlik, M. Moszynski, A. Nassalski et al., Investigation of absolute light output measurement techniques, IEEE Trans. Nucl. Sci., 54 (2007) 1367-1371.
18. J. T. M. de Haas and P. Dorenbos, Advances in yield calibration of scintillators, IEEE Trans. Nucl. Sci., 55 (2008) 1086-1092.
19. W. Drozdowski, A. J. Wojtowicz, S. M.Kaczmarek and M. Berkowski Scintillation yield of $Bi_4Ge_3O_{12}$ (BGO) pixelcrystals Physica B 405 (2010) 1647–1651.
20. D. Wahl, V. B. Mikhailik and H. Kraus, Monte-Carlo refractive index matching technique for determining the input parameters for simulation of the light collection in scintillating crystals, Nucl. Instr. Meth. Phys. Res. A 570 (2007) 529-535.
21. O. S. Kushnir, Effect of multiple reflections of light on the optical characteristics of crystals, J. Optics A. Pure Appl. Opt. 5 (2003) 478-488.
22. W. L. Bond, Measurements of the refractive indices of several crystals, J. appl. Phys., 36 (1965) 1674-1677.
23. H. Kraus, V. B. Mikhailik and D. Wahl, Multiple photon counting coincidence (MPCC) technique for scintillator characterisation and its application to studies of $CaWO_4$ and $ZnWO_4$ scintillators, Nucl. Instr. Meth. Phys. Res. A, 553 (2005) 522-534.
24. V. B. Mikhailik and H. Kraus, Development of techniques for characterisation of scintillation materials for cryogenic application, Rad. Measur. (2013) accepted, in press.
25. Radiant ZEMAX, http://www.radiantzemax.com/

26. G. Keil, Design principles of fluorescence radiation converters, Nucl. Instr. Meth., 89 (1970) 111-123.

27. F. Bauer, Corbeil, M. Schmand and D. Henseler, Measurements and ray-tracing simulations of light spread in LSO crystals, IEEE Trans. Nucl. Sci., 56 (2009) 2566-2573.

28. M. Moszyński, M. Balcerzyk, W. Czarnacki, et al., Characterization of $CaWO_4$ scintillator at room and liquid nitrogen temperatures, Nucl. Instr. Meth. Phys. Res. A 553 (2005) 578-591.

29. R. V. Vasiliev, S. B. Lubsandorzhiev, B. K. Lubsandorzhiev, et al. Measuring the lght yield in a $CaMoO_4$ scintillating crystal, Nucl. Instr. Meth. Phys. Res. A, 53 (2010) 795-799.